\documentclass[11pt,titlepage]{article}

\usepackage{setspace} 
\usepackage{amsmath}

\usepackage{graphicx}

\usepackage[round,numbers,sort&compress]{natbib} 
\title{Stochastic oscillations induced by intrinsic fluctuations in a self-repressing gene: a deterministic approach.}

\author{Jingkui Wang \\
	Laboratoire de Physique des Lasers, Atomes, et Mol\'ecules, CNRS,
        UMR8523 \\ 
	Universit\'e Lille 1, F-59655 
	Villeneuve d'Ascq, France\\
	\and Marc Lefranc \\
	Laboratoire de Physique des Lasers, Atomes, et Mol\'ecules, CNRS,
        UMR8523 \\ 
	Universit\'e Lille 1, F-59655 
	Villeneuve d'Ascq, France\\
        \and Quentin Thommen\thanks{
           Corresponding author.  Address: 
        quentin.thommen@univ-lille1.fr}\\
	Laboratoire de Physique des Lasers, Atomes, et Mol\'ecules, CNRS,
        UMR8523 \\ 
	Universit\'e Lille 1, F-59655 }

\date{}

\begin{document}

\maketitle

\abstract{ Biochemical reaction networks are subjected to large
  fluctuations attributable to small molecule numbers, yet underlie
  reliable biological functions. Most theoretical approaches describe
  them as purely deterministic or stochastic dynamical systems,
  depending on which point of view is favored. Here, we investigate
  the dynamics of a self-repressing gene using an intermediate
  approach based on a moment closure approximation of the master
  equation, which allows us to take into account the binary character
  of gene activity. We thereby obtain deterministic equations that
  describe how nonlinearity feeds back fluctuations into the
  mean-field equations, providing insight into the interplay of
  determinism and stochasticity. This allows us to identify regions of
  parameter space where fluctuations induce relatively regular
  oscillations.

}

\clearpage

\section*{Introduction}
Most cellular functions are controlled by molecular networks involving
genes and proteins which regulate each other so as to generate the
adequate dynamical behavior. A major goal of systems biology is to
understand how sophisticated functional modules emerge from the
combination of elementary processes such as transcriptional
regulation, complex degradation, active transport,... and how each of
these processes influences the collective dynamics~\cite{Hartwell99}.

A specificity of regulatory networks viewed as dynamical systems is
that they are both strongly nonlinear and inherently stochastic, which
considerably complicates the mathematical analysis. In a cell, protein
and mRNA molecules are often found in low abundance so that variations
of their copy numbers by one unit represent significant fluctuations.
Furthermore, DNA fragments carrying genes are single-copy molecules
which have only a few possible configurations depending on promoter
occupancy. When its transcription is regulated by a single protein, a
gene can essentially be in two states: free, or bound to its
transcription factor. Gene activity is then be described
mathematically by a binary variable, which more generally can also
account for the transcriptional pulsing that has been observed both in
prokaryotes~\cite{Golding05} and
eukaryotes~\cite{xavier07,Chubb06,Suter11:_mammalian_bursting,Harper11:_stochastic_cycles}.
The stochastic dynamics of the gene randomly flipping between the
bound and free states with probabilities depending on transcription
factor abundance is a major source of intrinsic fluctuations, all the
more at it was shown that this flipping can occur at time scales which
are comparable to other biochemical processes~\cite{Golding05}. While
stochasticity in gene networks has been often viewed as an undesirable
perturbation blurring deterministic behavior, it is increasingly
recognized that noise can in fact be harnessed so as to become a
functional component of a regulatory network and make its dynamics
richer~\cite{eldar10:_funct,Simpson15042003,warren:144904,lestas2008noise,Aquino12}.
It is thus important to understand how the deterministic and
stochastic aspects of cellular processes interact and contribute to
the same global dynamics, all the more as they are intimately coupled
in nonlinear systems.

However, even moderately complex regulatory networks resist
mathematical analysis and require formidable computational resources.
A natural strategy to study such general questions as the interplay of
dynamics and noise is to focus on small genetic networks comprising
only a few elementary components, whose analysis can identify the key
mechanisms and parameters and cast light on the dynamics of more
complex networks. This approach is all the more valuable as the recent
developments of synthetic biology allow experimental tests of the
theoretical analyses~\cite{Stricker08:robust_oscillator}.

Here, we study how stochastic fluctuations in gene activity feed back
into the deterministic dynamics of the smallest genetic network, which
consists of a single gene repressed by its own protein. This system is
an ideal workbench to investigate how the dynamics of the network
emerges from the properties of its elementary components. In fact,
this motif is very common in transcriptional networks and is thus
biologically relevant (around 40\% of \textit{Escherichia coli}
transcription factors are
self-repressing~\cite{Hermsen2010a,Salgado01012001,Keseler01012005}).
Self-repression is known to be an important ingredient for generating
oscillatory behavior~\cite{Novak_08}. For instance, Hirata \emph{et al}
proposed that the somite clock network is governed by the
self-repressing gene \emph{Hes1}~\cite{Hirata02a}. Accordingly, the
dynamics of the self-repressing gene has been actively investigated
throughout mathematical biology~\cite{Goodwin65a, Griffith68a,
  Bliss82a, Goldbeter95a, Leloup99a, Lewis2003a, Monk2003a,
  Jensen2003a,Morant09}.

Most theoretical analyses of the self-repressing gene based on a
deterministic description assume that gene state flipping occurs on
much faster time scales than other processes such as transcription,
translation, and degradation. The flipping dynamics can then be taken
into account through an average activity, which adapts to protein
concentration either instantaneously or after a time delay. If
intrinsic fluctuations are neglected, the analysis of the rate
equations reveals that oscillatory behavior can only be found by
either (1) introducing an explicit time delay in the equations (e.g.,
to take into account the transcriptional
dynamics~\cite{Lewis2003a,Monk2003a,Jensen2003a,Novak_08,tiana07:_oscil,Jensen_Current_Opinion});
(2) inducing an implicit time delay via a reactional step, which can
be intrinsic~\cite{Bliss82a} or describe transport between two
compartments~\cite{Leloup99a}; (3) incorporating complex degradation
mechanisms~\cite{Tyson99:_per_tim,Novak_08,Jensen_Current_Opinion,Morant09}.
However, recent experiments have shown that gene activity may display
an intrinsic dynamics on time scales comparable to that of other
cellular
processes~\cite{Golding05,Chubb06,Suter11:_mammalian_bursting,Harper11:_stochastic_cycles}.
This may be taken into account in a deterministic model by introducing
a average gene activity variable, which reacts gradually to protein
concentration~\cite{Francois05}. In particular, how such a
transcriptional delay and a nonlinear degradation mechanism conspire
to generate oscillations has been studied in detail by Morant \emph{et
  al.}~\cite{Morant09}, who obtained analytical expressions for the
instability thresholds.

To take into account the binary nature of gene state and its
stochasticity, the most general approach to study the dynamics of the
self-repressing gene is to use the chemical master equation
(CME)~\cite{VanKampen}. The steady-state solution of the CME provides
the probability distribution of molecular copy numbers, characterizing
both the averages and the fluctuations around them. An analytical
solution of the CME for the self-repressing gene can be obtained when
the mRNA variable is considered to be fast and can be eliminated
adiabatically~\cite{hornos05,grima2012}, but this assumption is
unrealistic for the \emph{Hes1} feedback network, where mRNA and
protein have similar lifetimes~\cite{Hirata02a}. A classical strategy
for approximating the CME is the system-size expansion also known as
Van Kampen's $\Omega$-expansion~\cite{VanKampen}. Assuming that the
system size is large but not infinite, the solution is expanded in
powers of the inverse system size. The deterministic mean-field
equations are obtained at lowest order while next-to-leading order
corrections determine finite-size fluctuations in the so-called linear
noise approximation (LNA). This approach can be used to estimate the
amplitude of fluctuations~\cite{elf03:_lna} but also to determine
their spectrum. In the latter case, the LNA has been useful to
characterize the appearance of stochastic oscillations in parameter
regions where the mean-field equations predict stable steady
behavior~\cite{mckane05:_predat,mckane07:_amplif} or to verify that
oscillations predicted by a deterministic modeling persist in presence
of fluctuations~\cite{Galla_09}, two problems which has been much
studied~\cite{loinger:051917,blossey:2008,Barrio_06,Lepzelter_10}. To
overcome the fact that this method does not allow one to determine
precisely when the steady state loses stability, Scott, Ingalls and
Hwa proposed an extension of the LNA which takes into account how
fluctuations modify the linearized dynamics around steady state and
allows one to study how bifurcation diagrams are modified by
noise~\cite{Scott07}. However, all these methods based on system size
expansion assume that fluctuations vanish in the infinite size limit,
without affecting the average values. This assumption clearly does not
hold when the gene state is a binary variable which fluctuates between
two discrete values, regardless of system size. A different approach
must then be taken.

\begin{figure}[htbp]
  \begin{center}
    \includegraphics*[width=5in]{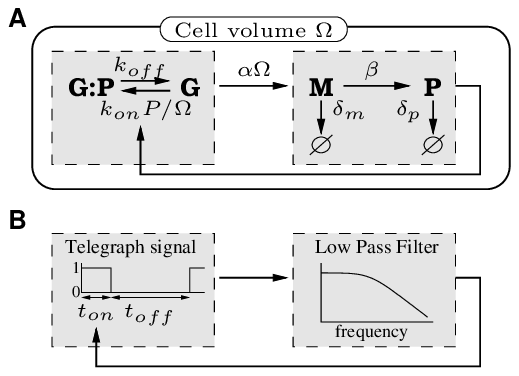}
  \caption[]{}
  \label{fig:fig1}
\end{center}
\end{figure}

In this paper, we derive a deterministic description of the stochastic
dynamics of a basic self-repressing gene circuit, with no
cooperativity in the transcriptional regulation and a linear
degradation mechanism (Fig.~\ref{fig:fig1}A). The gene switches
stochastically between the active and inactive state, so that this
circuit can be viewed as a random telegraph signal generator, whose
output is sent through a low-pass filter before being fed back to
itself (Fig.~\ref{fig:fig1}B). It is well known that a mean-field model
of this system is unconditionally stable (see, e.g., \cite{Morant09}).
Our main result is that an ODE model taking fluctuations into account
predicts oscillatory behavior in a region of parameter space where we
observe relatively regular spiking in protein concentration.

Using a moment closure approximation of the master
equation~\cite{gillespie09:_moment,lee09:closure,grima2012}, we derive
deterministic differential equations which generalize the usual
mean-field description while taking into account the binary nature of
the gene state variable. These equations describe the combined time
evolution of average quantities and of fluctuations around them, and
correctly predict the appearance of oscillations, as well as the
stationary state value of the network without any assumption on the
gene switch time scale nor on the statistical distribution of random
variables. We then explain the appearance of stochastic oscillations
by a simple resonance effect between the characteristic times scales
of the stochastic network and derive an analytical criterion for their
appearance. Finally, fitting this criterion with parameter values
relevant for the Hes1 network suggests that the mechanism we describe
may be exploited to generate robust oscillations in \emph{Hes1}
expression. Our findings highlight the functional role of intrinsic
fluctuations arising from the gene state-flip dynamics as an important
ingredient for shaping the dynamics of genetic networks.

\section*{Results and discussion}

\subsection*{Deterministic models taking fluctuations into account}

Three stochastic variables characterize the network dynamical state:
the gene state $g$, the mRNA copy number $m$ and the protein copy
number $p$. The time evolution of the probabilities $P_{g,m,p}$ of
being in a state with given values of $g$, $m$ and $p$ is given by the
following chemical master equation:
\begin{subequations}
  \label{eq:master}
\begin{eqnarray}
\frac{d}{dt}P_{g,m,p} & = & \left(-1\right)^g \left[ \frac{k_{on}}{\Omega} \;(p+1-g)\;P_{1,m,p+1-g}-k_{off} P_{0,m,p-g}\right]\\
&&  +\delta_{g,1}\alpha \Omega \left[P_{g,m-1,p}-P_{g,m,p} \right]
 +\beta m \left[P_{g,m,p-1}-P_{g,m,p} \right]\\
&&+\delta_m \left[(m+1)P_{g,m+1,p}-mP_{g,m,p}  \right]
+\delta_p \left[(p+1)\;P_{g,m,p+1}-p\;P_{g,m,p}  \right]
\end{eqnarray}
\end{subequations}
which can be read from (Fig.~\ref{fig:fig1}A) and provides the most
general description of the dynamics. The parameters $k_{\mathrm{on}}$
and $k_{\mathrm{off}}$ characterize the kinetics of protein-DNA
binding and unbinding, respectively. The transcription rate and
translation rate are $\alpha/\Omega$ and $\beta$, where $\Omega$ represents
the cell volume, and $\delta_m$ and $\delta_p$ are the mRNA and
protein degradation rates. The equations are normalized so that in the
large volume limit, the average gene activity $\langle g \rangle$ and
average concentrations $\langle m \rangle/\Omega$ and $\langle p
\rangle/\Omega$ become independent of $\Omega$.

Unfortunately, the master equation has generally no analytical
solution. Contrary to the mRNA and protein copy numbers, which become
much larger than one in the large volume limit and have then
negligible fluctuations when a single molecule is created or
destroyed, the gene state is a binary variable and its relative jump
size does not decrease. Therefore, the standard approximation method
based on the large-volume expansion of the master equation with the
van~Kampen ansatz fails~\cite{VanKampen}. Alternatively, the chemical
master equation can be reformulated as an infinite hierarchy of
coupled differential equations whose variables are the joint cumulants
of the random variables $g$, $m$, and $p$~\cite{VanKampen}. This
strategy leads to deterministic differential equations taking the
fluctuations into account and having the mean-field rate equations as
a limiting case.

To be specific, let us consider the equations describing the time
evolution of the averages of gene activity and mRNA and protein
concentrations in the infinite volume limit:
\begin{subequations}
\label{eq:Model0}
\begin{eqnarray}
 \frac{d}{dt}\langle \mathcal{P} \rangle& =&  \beta\,\langle \mathcal{M} \rangle-\delta_p \,\langle \mathcal{P} \rangle; \\ 
 \frac{d}{dt}\langle \mathcal{M} \rangle &=&  \alpha\,\langle g \rangle -\delta_m\,\langle \mathcal{M} \rangle; \\ 
 \frac{d}{dt}\langle g \rangle &=&  k_{off}\,\left[1-\langle g
   \rangle\right]-k_{on}\,\left[\langle g\, \mathcal{P} \rangle \right],\nonumber\\
 &=&  k_{off}\,\left[1-\langle g \rangle\right]-k_{on}\,\left[\langle g \rangle\langle \mathcal{P} \rangle+\textrm{cov}\left(g,\mathcal{P}\right) \right],\label{eq:geneaverage}
\end{eqnarray}
\end{subequations}
where $ \mathcal{M}$ (resp., $ \mathcal{P} $) denotes the
mRNA (resp., protein) concentration $ m /\Omega$ (resp.
$ p /\Omega$), $\langle x \rangle = \sum_{g,m,p} x
P_{g,m,p}$ is the average of the stochastic variable $x$ and
$\textrm{cov}\left(x,y\right)=\langle xy \rangle -\langle x \rangle
\langle y \rangle$ is the covariance of $x$ and $y$. These equations
are derived following the approach described in
the Supporting Material. Because of the nonlinear term
associated with DNA-protein binding in Eq.~\eqref{eq:geneaverage},
this equation can only be reformulated in terms of the average values
$\langle x\rangle$ by introducing the covariance term
$\textrm{cov}\left(g,P\right)$. This term does not appear in the usual
rate equations describing the kinetics of the self-repressing gene. It
describes the feedback from stochastic fluctuations into the dynamics
of the average values and plays therefore a key role to model the
influence of the gene state-flip dynamics. Eqs.~\eqref{eq:Model0} also
indicate that the dynamics of mRNA and proteins behaves as a low-pass
filter whose input is the mean gene state $\langle g \rangle$ and
output is the mean protein concentration $\langle P \rangle$. The
cut-off frequency of this low-pass filter depends only on mRNA and
protein degradation rates and has a typical value of
$\omega_c=\frac{\delta_m\delta_p}{\delta_m+\delta_p}$.

Eqs.~\eqref{eq:Model0} are only the first of an infinite hierarchy of
equations where time derivatives of the averages are expressed in
terms of averages and covariances, the time derivatives of covariances
are expressed in terms of covariances and third-order cumulants, and
so on (see Supporting Material). In order to truncate this infinite
hierarchy to a finite set of equations, a closure approximation must
be used. For instance, the usual rate equations are obtained when
infinite cell volume and vanishing covariances are assumed (i.e., the
$\textrm{cov}\left(g,P\right)$ term in Eqs.~\eqref{eq:Model0} is set
to $0$). The approximation neglects all fluctuations and assumes that
all variables have precise values, which conflicts with the binary
nature of the gene state. Here, we derive and analyze a higher-order
model, named thereafter the Third-Order Truncation (TOT) model, by
keeping only three averages and five covariances as dynamical
variables and neglecting all cumulants of order three and greater.
This assumes that variables are Gaussian-distributed and that their
dynamics can be described by their averages and
covariances~\cite{lafuerza10:_gauss}. We also assume infinite cell
volume, so that the only remaining fluctuations are those induced by
the gene flipping dynamics. In this limit, protein and mRNA copy
numbers are also infinite and thus their variation by one unit is
negligible, whereas the gene state is a binary variable, whose time
evolution is similar to a random telegraph signal.

The TOT model is most conveniently expressed in terms of eight
differential equations which specify the time evolution of the
averages and covariances of $g$, $p$ and of a new variable
$u=({\beta\; m + \delta_m \; p })/({\delta_p+\delta_m})$, after
suitable rescaling. More precisely, the equations comprising the TOT
model, whose derivation is detailed in the Supporting Material
, read
\begin{subequations}
\label{eq:Model_K30text}
\begin{eqnarray}
 \frac{d}{dT}P& =&  \eta\left(U-P\right); \\ 
 \frac{d}{dT}U &=&  \Lambda G-P; \\ 
 \frac{d}{dT}G &=&  \rho\left(1-G-GP-\Delta_{G,P}\right); \\ 
 \frac{d}{dT}\Delta_{G,U} &=& \Lambda G\left(1-G\right)-\Delta_{G,P}-\rho\left[G\Delta_{P,U}+\left(P+1\right)\Delta_{G,U}\right];\\
 \frac{d}{dT}\Delta_{G,P}& =&  \eta\left[\Delta_{G,U}-\Delta_{G,P}\right]  -\rho\left[G\Delta_{P,P}+\left(P+1\right)\Delta_{G,P}\right]\\
\frac{d}{dT}\Delta_{U,U} &=&  2\left[\Lambda \Delta_{G,U}-\Delta_{P,U} \right]; \\
\frac{d}{dT}\Delta_{P,P} &=&  2\eta\left(\Delta_{P,U}-\Delta_{P,P}\right);\\
\frac{d}{dT}\Delta_{P,U} &=&  \Lambda \Delta_{G,P}-\Delta_{P,P}+\eta \left[\Delta_{U,U}-\Delta_{P,U} \right].
\end{eqnarray}
\end{subequations}
where $P$, $U$, $G$ are the rescaled averages of the random variables
$p$, $u$ and $g$, and the $\Delta_{X,Y}$ are the corresponding
covariances. The three control parameters $\eta$, $\Lambda$ and $\rho$
are defined below.


\subsection*{The dynamics is controlled by three key parameters}
\label{sec:dynam-contr-three}

The biochemical reaction network of the self-repressing gene
(Fig.~\ref{fig:fig1}-A) has six independent kinetic parameters. Three
parameter combinations represent scales and thus can be taken out of
the equations by rescaling time as well as mRNA and protein
concentrations. These are : $\frac{k_{off}}{k_{on}}$, which is the
protein concentration at which gene is half repressed; $\frac{\delta_p
  k_{off}}{\beta k_{on}}$ which is the mRNA concentration
corresponding to half-repression in steady state; and
$\frac{\delta_m+\delta_p}{\delta_m\delta_p}$, which is the response
time of the low pass filter. There remain three reduced parameters,
denoted below by ${\rho,\Lambda,\eta}$, which control the dynamics and
are discussed below.

The parameter
$\rho=k_{off}\frac{\left(\delta_m+\delta_p\right)}{\delta_p\delta_m}$
measures the gene unbinding rate relative to the cut-off frequency of
the low-pass filter. A low value of $\rho$ indicates that the low-pass
filter transmits all the fluctuations of the gene state : the protein
concentration time profile displays squared waveforms enslaved to the
gene flip. By contrast, a high value of $\rho$ corresponds to the case
where the low-pass filter averages out the gene flip dynamics :
protein concentration evolves with small amplitude fluctuations around
its mean value.

The second reduced parameter $\Lambda=\frac{\alpha\beta
  k_{on}}{\delta_m\delta_p k_{off}}$ corresponds to the maximum
possible protein concentration relative to the half-repression protein
concentration threshold $\frac{k_{off}}{k_{on}}$. Dynamically,
$\Lambda$ characterizes the amplification of the gene telegraph signal
sent to the low pass-filter. A low value of $\Lambda$ ($\Lambda\ll 1$)
indicates that the gene remains unbound most of the time; the average
period of one gene on/off cycle is essentially the ``on'' state
duration $t_{on}$. On the contrary, a high value of $\Lambda$
($\Lambda \gg 1$) indicates that the gene is repressed most of the
time; the period of the gene on/off cycle is essentially the ``off''
state duration $t_{off}$, governed by $k_{off}$. Thus, $\Lambda$ can
also be viewed as characterizing the strength of the feedback from the
gene to itself via its protein.

The third parameter
$\eta=\frac{\left(\delta_m+\delta_p\right)^2}{\delta_p\delta_m}$
characterizes whether the protein and mRNA degradation rates are
balanced or not. This indicator reaches a minimum value of $4$ for
equal degradation rates ($\delta_m=\delta_p$) and goes to infinity as
one of the degradation rates becomes negligible compared to the other.
It is worth noting that the expressions of all key parameters $\rho$,
$\Lambda$, and $\eta$ are symmetric with respect to exchange of
$\delta_m$ and $\delta_p$. As a consequence, the dynamical properties
are unchanged if the mRNA and protein degradation rates are swapped, a
fact which was already noted in~\cite{Morant09}. To distinguish the
two regimes which have identical $\rho$, $\Lambda$ and $\eta$
parameter values but different values of $\delta_m$ and $\delta_p$, we
will later consider the ratio of protein and mRNA degradation rates
$\delta=\delta_p/\delta_m$, with
$\eta=\left(1+\delta\right)^2/\delta$. Obviously, the value of $\eta$
is unchanged under the transformation $\delta \leftrightarrow
1/\delta$.


\subsection*{The deterministic TOT model reproduces well the time
  averages of the stochastic dynamics}

To assess the influence of stochastic fluctuations of gene state on
the dynamics of the self-repressing gene, we first performed
stochastic numerical simulations to determine the values of the time
averages and covariances of the rescaled random variables $G$, $M$,
and $P$ (see Methods) as a function of the control parameters.
These time averages were then compared to the fixed point values of
two truncations of the cumulant equation hierarchy: the
rate equation model, defined by Eqs.~\eqref{eq:Model0} with the
covariance term set to zero, and the TOT model defined by
Eqs.~\eqref{eq:Model_K30text}, where the third-order cumulants have been
set to zero. These models are defined by sets of ordinary differential
equations, whose fixed points are specified by the values of the
variables such that all time derivatives are zero. These fixed points
are usually stable and thus reflect the stationary regime, however we
shall see later that they may become unstable in some conditions,
indicating the appearance of spontaneous oscillations.

Let us first examine how the average gene activity depends on $\rho$
(which characterizes the gene response time scale) and $\Lambda$
(which characterize feedback strength) when protein and mRNA lifetimes
are identical ($\delta=1$). Gene average activity  determined by
stochastic simulations is shown in Fig.~\ref{fig:fig2}-A. The
rate equation model correctly predicts the output of these stochastic
simulations only when gene dynamics is fast ($\rho\rightarrow\infty$)
or when gene repression is small ($\Lambda\ll 1$)
(Fig.~\ref{fig:fig2}-B). In contrast to this, the TOT model
predicts quantitatively gene average activity in the entire
$\left(\rho,\Lambda\right)$ plane (Fig.~\ref{fig:fig2}-C), in
particular in regions where the rate equation approximation fails.

A more detailed assessment of the TOT model accuracy is provided in
Fig.~\ref{fig:fig3}, which shows how the time averages and
first joint cumulants of $G$, $M$, and $P$ evolve with $\rho$ and
$\delta$, depending on whether they are computed from stochastic
simulations (Fig.~\ref{fig:fig3}, left column) or from the
fixed point values of the TOT model (Fig.~\ref{fig:fig3}, right
column). The computations are carried out in the strong feedback (high
repression) limit ($\Lambda=100$). Note that in the rate equation
approximation, all averages would be constant and the covariances
would vanish. Fig.~\ref{fig:fig3} displays only a subset of
components of the TOT model fixed point, from which the other can be
obtained with the relations $M^*=P^* = \Lambda G^*$,
$\Delta_{MM}^*=\Lambda\Delta_{GM}^*$ and
$\Delta_{PM}^*=\Delta_{PP}^*$. An important finding is that $\rho$ is
the main parameter controlling the averages, as the curves obtained
for various values of $\delta$ superimpose remarkably well
(Fig.~\ref{fig:fig3}-A, left column). This remains
approximately true for the covariances and higher-order cumulants,
although to a variable extent.

Fig.~\ref{fig:fig3} shows that the fixed point values of the
TOT model are in good quantitative agreement with the numerical
estimators. Concerning the averages (first order cumulants), the
overall shapes of the curves, with a maximum around $\rho=1$, are very
similar and the evolution of this maximum with $\delta$ is reproduced
(Fig.~\ref{fig:fig3}-A). The main discrepancy is that the
transition from the fast to the slow gene regime is more abrupt in the
TOT model than in stochastic simulations, presumably because
higher-order contributions to the averages are neglected. The global
evolution of the covariances is also well reproduced, and the values
of $\rho$ where $\Delta_{GP}$ becomes zero are also well predicted for
the different values of $\delta$ (Fig.~\ref{fig:fig3}-B).
Similarly, the variation of $\Delta_{GM}$ with $\delta$ is captured.
(Fig.~\ref{fig:fig3}-C). However, the TOT model fixed point
values overestimate the covariances $\Delta_{GM}$ and $\Delta_{PP}$
(Fig.~\ref{fig:fig3}-C,D) in the fast gene limit. Still, the
asymptotic values of the TOT model steady states (summarized in
Table~S1 in the Supporting Material) are correctly obtained.

A key assumption of the TOT model is that the third-order joint
cumulants $K_{G,M,P}$ and $K_{G,P,P}$ vanish. However,
Figs.~\ref{fig:fig3}-E,F show that they take rather large
values in the stochastic simulations, of the order of $\Lambda$, in
the slow gene limit. One may thus wonder why the TOT model is so
effective in this regime. A careful analysis of the structure of the
equations (Eq.~S8 in the Supporting Material)
solves this paradox. The key point is
that in the expressions of the time derivatives of the covariances,
the third-order joint cumulant are weighted by $\rho$ so that their
dynamical influence vanishes in the slow-gene limit (see
Sec.~S4 in the Supporting Material). Therefore, the TOT model is a valid
approximation for both fast and slow gene dynamics, and provides a
reasonable description of the dynamics in the intermediate regime.

Simple dynamical considerations can explain the observed dependency of
cumulants on $\rho$. If $\rho\gg 1$, the gene flip dynamics is
averaged by the low pass filter, as previously mentioned, and the
stationary regime is correctly predicted by the fixed point values of
the rate equations. In this limit, the gene remains bound or unbound
for very short amounts of time, during which mRNA and proteins copy
numbers can be considered as constant. RNA and protein levels keep a
memory of many previous state switching cycles, and reach a stationary
state with a probability distribution which is expected to be
Gaussian, as is confirmed by the vanishing of third order cumulants
$K_{G,M,P}$ and $K_{G,P,P}$. The coefficient of variation
$CV=\frac{\sqrt{\Delta_{PP}^*}}{P^*}$ tends to zero as $\rho$
increased, indicating that fluctuations in protein concentration
become negligible compared to the average concentration in the limit
of fast gene dynamics ($\rho\rightarrow\infty$). The negative value
taken by the covariance $\Delta_{GP}$ is consistent with the
negative feedback loop structure of the genetic network.

Conversely, if $\rho\ll 1$, the gene reacts infinitely slowly. The
dynamics is then driven by the gene jumping between two states
according to a Poisson process. During the time where the gene is
active (resp. inactive), protein and mRNA levels quickly converge to
their maximum value $\Lambda$ (resp. to zero); at the end of an gene
switching state cycle, variables are always in the same state with no
memory of previous cycles. This is consistent with the positive value
of $\Delta_{GP}$. Protein concentration temporal profiles feature a
sequence of squared shape spikes, distributed in time according to a
Poisson process, and characterized by a coefficient of variation
$CV\approx\sqrt{\Lambda}$ increasing with the overall production rate
$\Lambda$. Thus, fluctuations are enhanced by a slow gene and a high
repression.

Then, a natural question is whether there exists between these two
limit cases a dynamical regime which both behaves deterministically,
as in the fast-gene limit, and displays strong variations of the
protein concentration, as in the slow-gene limit. Such dynamical
behavior would feature a sequence of protein concentration spikes, but
with a time interval distribution more regular than a Poisson process.
This intuition is based on the fact that when the gene flip frequency
and the cut-off frequency of the low-pass filter are resonant ($\rho
\approx 1$), the random fluctuations of gene flips generated by the
Poisson process should be partially buffered by the low-pass filter.
This mechanism should prevent spike bunching, generating a more
regular dynamical behavior, which would be the stochastic analogue of
an oscillatory behavior (that we termed thereafter stochastic
oscillations). To assess the veracity of this idea, we developed a
criterion to quantify the regularity of stochastic oscillations,
described in next section.

\subsection*{Negative feedback induces protein spike antibunching}
\label{sec:negat-feedb-induc}

The regularity of a stochastic oscillatory behavior is often
quantified using a temporal autocorrelation
function~\cite{Simpson15042003,Galla_09,coulon2010}. This measure is
sensitive to reproducibility both in time and in amplitude. However,
temporal regularity is certainly more relevant than amplitude
regularity for biological protein signals. The highly nonlinear
response of many signaling cascades can protect them against
fluctuations in amplitude, for example by saturating output above an
input threshold. A standard technique for assessing temporal
regularity is to divide the state space into two regions I and II and
to study the distribution of the times where the system leaves I to
enter II. It is often useful to require a minimal excursion in region
II to avoid spurious transitions induced by noise. Here, we detect
events where the protein level crosses successively the mean protein
level $P^*$ and the $P^{'*}=P^*+0.25\sqrt{\Delta_{pp}^*}$ level before
falling back below the mean protein level.

Given the list of times where the system transits from low to high
protein levels, we compute the probability of detecting $n$
transitions within a time interval of fixed duration. To be specific,
we select a time interval equal to ten times the average time between
two events, and characterize the probability distribution of the
number of events by the variance to mean ratio, also known as the Fano
factor~\cite{Fano47}. This method is inspired by how the temporal
distribution of photons from a light source is generally
characterized, with the event of interest being a photon detection. A
Fano factor close to unity is obtained when time intervals between
events follow a Poisson distribution. A Fano factor greater (less)
than unity indicates super-Poissonian (resp., sub-Poissonian) behavior
corresponding to a bunching (resp., anti-bunching) of protein spikes.
Spike anti-bunching can be viewed as a stochastic counterpart of
deterministic oscillations. While using the coefficient of variation
of the interspike interval would give similar results, the method
described above has the advantage to take into account correlations
between the successive transitions.

Figure~\ref{fig:fig3} displays stochastic simulations of the
chemical reaction network of Fig.~\ref{fig:fig1} for a slow, an
intermediate and a fast gene, as well as the probability distribution
of the number $n$ of transitions within a given time window. As
expected, protein spikes in the slow gene case
(Fig.~\ref{fig:fig3}-A) are slaved to the switching process,
leading to a Poisson probability distribution for $n$
(Fig.~\ref{fig:fig3}-D) and accordingly a unity Fano factor. In
the intermediate gene response time case
(Fig.~\ref{fig:fig3}-B), protein spikes are mostly antibunched
(see black circles). The probability distribution of spike number is
gaussian-like (Fig.~\ref{fig:fig3}-E), the Fano factor being
around 0.25. This anti-bunching degrades in the case of a fast gene
(Fig.~\ref{fig:fig3}-C) with the Fano factor rising to 0.9.
Thus, we observe a resonance effect which results from the coincidence
of the gene response time to protein variations and the time during
which previous gene state history is remembered, which is controlled
by the protein and mRNA decay rates.

We studied systematically how the Fano Factor depends on the gene
resonance parameter $\rho$ and the relative protein decay rate
$\delta$ in stochastic simulations (Fig.~\ref{fig:fig5}). We
found that the regularity of protein spikes is reinforced by (1)
similar mRNA and protein decay rates ($\delta\approx 1$), (2) a
resonance parameter close to unity ($\rho\approx 1$), and (3) a
sufficiently strong feedback ($\Lambda>1$), as shown in
Fig.~\ref{fig:fig5}-A,B. Thus, the most regular oscillations
are observed when the gene cycling period resonates with the average
mRNA/protein lifetime.

The lack of symmetry with respect to the transformation $\delta
\leftrightarrow 1/\delta$ for low values of $\Lambda$
(Fig.~\ref{fig:fig5}-B) results from numerical difficulties to
reach the infinite cell volume limit for small $\delta_p$ ($\delta\ll
1$). As a control, we checked that the Fano Factor is almost
independent of the ratio $\beta/\delta_p$ (see
Fig.~\ref{fig:fig5}-C ), which determines the protein to mRNA
ratio. This confirms that the reduced parameter $\Lambda$ is the
relevant parameter for describing the overall production rate and thus
the amplification factor of the feedback loop.

In the large $\Lambda$ limit, it is expected that the gene spends most
of the time in the ``off'' state so that the average duration of one
``on''/``off'' cycle is approximately given by
$\tau_{\mathrm{off}}=1/k_{\mathrm{off}}$ in original time units. To
study the interplay between the gene state dynamics and the protein
spike dynamics, a useful indicator is $\chi=k_{off}
\langle T_s \rangle$ where $T_s$ is the average time interval between
two spikes. In the slow gene limit ($\rho\rightarrow 0$), $\chi$ tends
to 1, indicating that protein and mRNA are slaved to the gene dynamics
in a ``fire and degrade'' mode. Conversely, the high value of $\chi$
in the fast gene limit indicates that the gene dynamics is too fast to
be relevant and justifies an adiabatic elimination of the gene state
variable. In the parameter region where spikes are more regular, the
intermediate values taken by $\chi$ (between 1 and 10) reveals that
the gene dynamics and the mRNA/protein dynamics influence each other
and generate together the stochastic oscillations observed.

\subsection*{A deterministic model predicts the appearance of
  stochastic oscillations}
\label{sec:determ-model-oscillations}


If a moment-closure model such as the TOT model
(Eqs.~\eqref{eq:Model_K30text}) is relevant to the dynamics of the
self-repressing gene, it should be able to predict the stochastic
oscillations evidenced in the previous section. While such models take
fluctuations into account, they are deterministic ODE models, where
the natural counterpart of the regular spiking observed in stochastic
simulations is the occurrence of self-sustained oscillations. A linear
stability analysis of the TOT model should then provide analytical
insight into the key parameters controlling the stochastic
oscillations.

Because the TOT-model is eight-dimensional, and thus difficult to
analyze analytically, we considered a reduced version of it, obtained
under the assumption that $\rho \ll \eta,1$, which corresponds to the
slow-gene limit. In this limit, the following subset of equations
uncouples from the others, regardless of whether third-order joint
cumulants vanish or not (they can be obtained by setting $\rho=0$ in
Eqs.~\eqref{eq:Model_K30text}):
\begin{subequations}
  \label{eq:Mod_red}
  \begin{eqnarray}
 \frac{d}{dT}P& =&  \eta\left(U-P\right); \\ 
 \frac{d}{dT}U &=&  \Lambda G-P; \\ 
 \frac{d}{dT}G &=&  \rho\left(1-G-GP-\Delta_{G,P}\right); \\ 
 \frac{d}{dT}\Delta_{G,U} &=& \Lambda G\left(1-G\right)-\Delta_{G,P};\\
 \frac{d}{dT}\Delta_{G,P}& =&  \eta\left[\Delta_{G,U}-\Delta_{G,P}\right],
  \end{eqnarray}
\end{subequations}
Besides the averages, the reduced model~\eqref{eq:Mod_red}
incorporates the dynamics of the covariances of the gene state
variable with protein and mRNA levels. These are indeed the variables
that most directly capture the influence of gene state fluctuations on
the dynamics of the self-repressing gene.

Remarkably, a stability analysis of Eqs.~(\ref{eq:Mod_red}) reveals
that this system exhibits a Hopf bifurcation leading to oscillatory
behaviour when the Routh--Hurwitz oscillation criterion
\begin{equation}
  \label{eq:RH_1}
  \mathcal{H}(\rho,\eta,\Lambda)=\rho^{2}\left(\frac{2\Lambda+1}{\Lambda+1}\right)^{2}+\rho\left[\frac{2\eta\Lambda+\eta-\Lambda^{2}}{\Lambda+1}\right]+\eta\;<0.
\end{equation}
is satisfied. Since the mean-field model of the self-repressing gene
is unconditionally stable, this shows that fluctuations can play a
functional role to promote oscillations. The expression of
criterion~\eqref{eq:RH_1} shows that $\Lambda$ and $\eta$ are the key
parameters controlling the appearance of oscillations in
Eqs.~\eqref{eq:Mod_red} and accordingly of regular spiking in the
stochastic simulations. In particular, it can be seen that the only
negative term in~\eqref{eq:RH_1} is $-\rho\Lambda^2/(\Lambda+1)$ and
thus that a large value of $\Lambda$ generally favors oscillations. On
the other hand a small value of $\eta$ is required for the
criterion~\eqref{eq:RH_1} to become negative over some range of
$\rho$. More precisely, it is easily shown that the existence of an
oscillation region requires that $\Lambda \ge 2\eta+4\sqrt{\eta}$
(when this inequality is satisfied, the assumption $\Lambda \gg 1$
under which it is derived holds naturally). Whether oscillations
actually occur, however, depends on the value of the resonance
parameter $\rho$.

When $\rho\rightarrow 0$ (i.e., the gene cycling period is much longer
than average protein/mRNA lifetimes), the criterion is never satisfied
because $\mathcal{H}=\eta>0$, and no oscillations occur. When
$\rho\rightarrow \infty$, the dominant term is obviously positive, and
no oscillations occur either. For intermediate values of $\rho$, the
quantity $\mathcal{H}$ becomes negative for sufficiently large
$\Lambda$ and for $\eta$ sufficiently close to its minimum value of
$4$, as discussed above. Thus, oscillations are favored when feedback
is strong and when the mRNA and protein degradation rates are
balanced. In the limit of large $\Lambda$ ($\Lambda\gg \eta$), the
criterion~\eqref{eq:RH_1} simplifies to
$\mathcal{H}(\rho,\eta,\Lambda)=4\rho^2-\rho\Lambda+\eta<0$ and
oscillations are then found for
$\rho\in[\rho_l,\rho_h]=[\eta/\Lambda,\Lambda/8]$, a rather wide
interval around $\rho\sim 1$. Note that since $\Lambda\gg \eta$, the lower
bound of the oscillation interval is naturally located in the region
$\rho \ll 1$ where the reduced model is valid. However, whether
oscillations actually occur for larger values of $\rho$, including the
upper bound $\Lambda/8$ is less clear.

Comparing with numerical simulations, we can see in
Fig.~\ref{fig:fig5}-A,B that the criterion~\eqref{eq:RH_1} is
fairly well satisfied in the region of parameter space where regular
stochastic oscillations are observed, especially when the influence of
the parameters $\rho$ and $\delta$ is considered. On the other hand,
while the influence of feedback strength $\Lambda$ on the appearance
of oscillations depending on degradation rate balance is correctly
captured, the criterion~\eqref{eq:RH_1} overestimates the value of
$\Lambda$ where regular oscillations are first observed
(Fig.~\ref{fig:fig5}-B). The good agreement observed over a
wide range of values of $\rho$ is all the more surprising as the
reduced model~\eqref{eq:Mod_red} is in principle valid only in the
slow gene limit (i.e., for $\rho$ sufficiently small). Indeed, it does
not predict correctly the average values when $\rho$ is of order 1 or
larger.

To conclude, the reduced model~\eqref{eq:Mod_red} captures well how
the mean-field variables and fluctuations interact to generate
relatively regular stochastic oscillations, although it does not
reproduce satisfactorily the average gene activity over the entire
parameter space. This suggests that the joint cumulants involving the
gene state are the most dynamically important ones, which is
consistent with the fact that the gene state remains a binary variable
in all limiting cases and thus is the most stochastic variable.

\subsection*{Oscillations in Hes1 expression match the criterion for
  fluctuation-induced oscillations}

The main result of this work is that stochastic fluctuations in a
self-repressing gene can play a functional role in promoting the
appearance of relatively regular oscillations in specific regions of
the parameter space. It is then natural to ask whether oscillating
self-repressing gene circuits found in biological systems operate in
the parameter region we have identified. One such circuit that has
been intensively studied is the \emph{Hes1} gene, which is believed to
be at the core of the somite clock~\cite{Hirata02a}. It is well known
that a crucial ingredient of oscillations in \emph{Hes1} expression is
the presence of a time delay, associated to transcription, translation
or transport. This time delay is often modeled as an explicit time
delay~\cite{Monk2003a,Lewis2003a,Jensen2003a,Barrio_06}, however it
can also result from a slow reactional
step~\cite{Goldbeter95a,Leloup99a,Morant09}.

In our system, the time delay is due to the finite gene response time
related to the binding/unbinding dynamics. This finite gene response
time can also be viewed as taking into account phenomenologically
other sources of delay, if they arise from reactional steps and thus
are exponentially distributed. More precisely, the gene can persist in
the ``off'' state for some time after protein level goes down because
of the characteristic time $\tau_g=k_{off}^{-1}$ (in original time
units) needed to switch from the ``off'' to the ``on'' state.
Therefore, this characteristic time can be viewed as the delay
inducing oscillations, and large-scale variations of protein
concentration will typically appear when it is not too small compared
to protein half-life. We found that these variations are more regular
when these two time scales are equal. Of course, the oscillations in
our model remain less regular than those observed in \emph{Hes1}
because, (1) the delay is exponentially distributed rather than
constant and (2) there is no cooperativity. Yet, the models are
sufficiently similar that if there is a parameter region where
fluctuations promote oscillations in our stochastic self-repressing
gene model, this should remain true for the \emph{Hes1} circuit since
oscillations would then be more robust to random variations of the
delay. Such random variations could be due for instance to the
presence of reactional steps. We should then expect this specific
parameter region to be selected by evolution.

A first interesting observation is that in the \emph{Hes1} oscillator,
the protein and mRNA half-lives are approximately equal, with reported
values of 22 and 24 minutes, respectively~\cite{Hirata02a}. This is
fully consistent with both our observation that regular oscillations
occur preferably for $\delta=1$ (Fig.~\ref{fig:fig5}) and the
fact that the oscillation criterion \eqref{eq:RH_1} for the 5-variable
reduced model \eqref{eq:Mod_red} indicates that $\eta$ should be as
close as possible to its minimum, which corresponds to balanced
half-lives. Note that this contrasts markedly with what is known for
the mean-field model, where making degradation rates unbalanced while
keeping their sum constant favors oscillations~\cite{Morant09}.

A crucial parameter for the regularity of the stochastic oscillations
is the resonance parameter $\rho$, which depends on the time delay and
on the mRNA and protein half-lives. However, the time delay in the
\emph{Hes1 circuit} is not known experimentally. In theoretical
investigations (see, e.g., ~\cite{Barrio_06,bernard06:_hes1}), it is
generally assumed that the time delay ranges from 10 to 40 minutes. We
assume here a value of 30 minutes, which is consistent with the fact
that for large half-lives, the oscillation period of 120 minutes is
approximately equal to four times the delay according to
Lewis~\cite{Lewis2003a}. With the known values for the mRNA and
protein half-lives (which translate to $\delta_p\sim\delta_m\sim
0.3$), this yields $\rho\sim 2$. Together with $\delta=1$, this value
corresponds precisely to the region of regular oscillations in
Fig.~\ref{fig:fig5}A. Furthermore, note that
Fig.~\ref{fig:fig5}D indicates that for $\rho=2$, the ratio
of the oscillation period to the delay $1/k_{off}$ is indeed close to 4.

Finally, other theoretical investigations (see, e.g.,
~\cite{Barrio_06,bernard06:_hes1}) assume that $\Lambda \gg 1$. This
in fact a natural condition, which simply requires that the maximum
protein level reached when the gene is fully active must be much
larger than the half-repression threshold. This ensures that the
protein level can go below and above this threshold in the course of
oscillations, and that the gene is strongly repressed when protein
level is high.

Taken together, these facts strongly suggest that the \emph{Hes1} mRNA
and protein half-lives have been tuned to be both close to the time
delay in order to make oscillations in \emph{Hes1} expression
robust against stochastic fluctuations in delay.

\section*{Conclusion}

In this paper, we have studied the stochastic dynamics of a simple
self-repressing gene model, in which the gene switches randomly
between the active and inactive state with a characteristic time which
can be arbitrarily small or large compared to mRNA and protein
lifetimes. The regularity of the protein spikes generated by the
dynamics was characterized using a Fano-like indicator. This allowed
us to evidence a dynamical resonance phenomenon, namely that a more
regular time evolution of protein concentration is observed for
certain values of the protein to mRNA degradation rate ratio and of
the gene response time. It should be stressed that there is neither
cooperativity nor nonlinear degradation in our model, so that the
regularity of the oscillations displayed in Fig.~\ref{fig:fig4}B could
easily be improved by using these two ingredients, as is done in most
theoretical investigations, or by considering a fixed time delay in
addition to the exponentially-distributed gene response time.

To understand the resonance phenomenon, we developed a deterministic
ODE model using a moment-closure approximation of the master equation.
This model describes the combined time evolution of the averages and
covariances of protein, mRNA and gene activity. Thus, it allows us to
describe how nonlinearity injects fluctuations into the average
dynamics, which can be substantially modified. The steady state of
this model predicts well how averages and covariances vary with the
gene response time and the ratio of mRNA and protein degradation
rates. In particular, it reproduces the fact that the average gene
activity is significantly reduced in the slow gene limit. In this
limit, a 5-dimensional model can be obtained, which incorporates the
three averages and the two covariances of the gene state with the
protein and mRNA concentrations. This model displays a Hopf
bifurcation which is not present in the mean-field model. Remarkably,
the parameter region where the reduced model oscillates matches the
region where the protein spikes are more regularly spaced, even though
the model predicts correct steady-state values only in the slow gene
limit. Therefore, deriving deterministic equations through a
moment-closure approximation of the master equation appears to be an
effective approach to describe the bifurcation diagram of stochastic
dynamical systems, which is generally a difficult problem (see,
e.g.,~\cite{Bratsun05:delay_stochastic_oscillations,Song10:_stochastic_bifurcation}).
This approach is all the more interesting as computer software is
available to derive the hierarchy of equations for the cumulants of
increasing order~\cite{gillespie09:_moment,Vidal2012}. The approach
describe here is well fitted to problems where one variable remains
microscopic, such as gene state, and where fluctuations dramatically
affect the average values. It thus brings a distinctive advantage
compared to other methods based on the linear noise
approximation~\cite{Scott07}.

To check whether the resonance effect discussed here is relevant in
real genetic oscillators, we examined the timescales reported for the
\emph{Hes1} self-repressing
gene~\cite{Hirata02a,Lewis2003a,Monk2003a}. In this circuit, the mRNA
and protein lifetimes are approximately equal to the time delay. We
found that this situation is characterized by the reduced parameters
$\rho=2$, $\delta=1$, which correspond precisely to the center of the
parameter region where regular protein spiking is observed. This
strongly suggests that the phenomenon of stochastic resonance we have
unveiled plays an important role for generating robust genetic
oscillations, independently of other oscillation-enhancing effects
such as cooperativity in the transcriptional
regulation~\cite{Purcell06112010} or nonlinear
degradation~\cite{Tyson99:_per_tim,Morant09}, which can be
simultaneously harnessed. A possibly related observation by Murugan
and Kreiman is that protein response times fluctuate less when mRNA
and protein lifetimes are closer~\cite{Murugan11}. More generally, we
believe that our findings provide a remarkable example of how
stochastic fluctuations, which are unavoidable in genetic networks,
may play a functional role to shape their
dynamics~\cite{eldar10:_funct}.

\section*{Methods}

To assess the validity of the various truncation schemes, we performed
numerical stochastic simulations of the chemical
network~(Fig.~\ref{fig:fig1}-A) for various values of $\rho$, $\Lambda$,
and $\delta$. To enforce a one-to-one relationship between the
original parameter space
$\{k_{on},k_{off},\beta,\alpha,\delta_m,\delta_p\}$ and the reduced
parameter space $\{\rho,\Lambda,\eta\}$, three constraints are
required. So we fixed (1) the ratio $\beta/\delta_p=10$ to enforce a
protein concentration 10 times bigger than mRNA concentration, which
is a realistic assumption for a biological network, (2) the repression
threshold of the gene $\Omega k_{off}/k_{on}=100$ to be consistent
with the assumption of infinite cell volume while keeping
computational time within reasonable limits, (3) $\delta_m=1$ to set
the time scale to the mRNA half-life. Since stochastic simulations
deal with copy numbers instead of concentration, the cell volume has
no influence and we fixed $\Omega=1$. The validity of the truncation
schemes investigated can then be assessed by comparing the values of
the averages in the stochastic simulation with the fixed point values
of the ODE models obtained by truncating the cumulant expansion.

The stochastic simulations were performed using an implementation of
the next reaction method (Gibson-Bruck
algorithm~\cite{GibsonBruck00}). The time interval used for the
numerical estimation of joint cumulants was chosen to ensure a
relative error of the estimator of the mean of the gene state smaller
than $10^{-4}$ by monitoring the convergence of the estimator and its
fluctuations. To compute an numerical estimation of the Fano factor,
we recorded 4000 interspike intervals after a transient whose duration
was chosen by the monitoring the convergence of the estimator for the
average.

\section*{Acknowledgments}
   This work has been supported by Ministry of Higher Education and
   Research, Nord-Pas de Calais Regional Council and FEDER through the
   Contrat de Projets \'Etat-R\'egion (CPER) 2007--2013, as well as by
   LABEX CEMPI (ANR-11-LABX-0007). The authors thank Benjamin Pfeuty
   for careful reading of the manuscript.

\newpage


\newpage

\section*{Figure Legends}

\subsubsection*{Figure~\ref{fig:fig1}.}
  \textbf{Schematic view of the self-repressing gene
      network.} (A) Biochemical reactions composing the network. $P$,
    $M$, $G$, and $G:P$ denote protein, mRNA, free gene and bound gene
    chemical species, respectively. The kinetic constants of the
    reactions are indicated, with $\Omega$ denoting cell volume. In
    the limit where $\Omega$ is large, the mRNA and protein copy
    numbers become macroscopic variables, with decreasing
    fluctuations, while the gene state remains microscopic and
    displays full-scale variations. (B) Block diagram representation
    of the network, consisting of a random telegraph signal generator
    representing the gene state-flip dynamics, and of a low-pass
    filter of cut-off frequency $\omega_c$ representing proteins and
    mRNA dynamics. The telegraph signal regulates its frequency and
    duty cycle through feedback from the low pass filter.

\subsubsection*{Figure~\ref{fig:fig2}.}
\textbf{Average gene activities as a function of the
      $\Lambda$ and $\rho$ parameter} (A) Numerical estimation of
    $\langle g \rangle$ using stochastic simulations with parameter
    values $k_{off}/k_{on}=100$, $\delta=1$, $\beta/\delta_p=10$. (B)
    Average gene activity predicted by rate equation (C) Fixed
    point value of gene activity in model~(\ref{eq:Model_K30text}). 

\subsubsection*{Figure~\ref{fig:fig3}.}
\textbf{Comparison of statistical quantities such as
      averages, covariants and higher-order cumulants obtained from
      stochastic simulations (left column) and from the TOT model
      (right column) as functions of parameter $\mathbf{\rho}$, for
      various values of $\mathbf{\delta}$.} (A) Average $G$; (B), (C)
    and (D) covariances $\Delta_{GP}$, $\Delta_{GM}$ and
    $\Delta_{PP}$; (E), and (F) third-order joint cumulants $K_{GMP}$
    and $K_{GPP}$. Curves for different values of $\delta$ are
    color-coded according to legend box. The value of $\Lambda=100$
    used in the simulations corresponds to strong feedback (strong
    gene repression). Stochastic simulations are performed while
    constraining $k_{off}/k_{on}=100$ and $\beta/\delta_p=10$ (see
    text for details). In each panel, thick lines (resp., thin) lines
    indicate positive (resp., negative) values.

\subsubsection*{Figure~\ref{fig:fig4}.}
\textbf{Protein spike antibunching.} (A,B,C). Time
    evolution of protein copy number for $\Lambda=100$, $\delta=1$ and
    $\rho=10^{-3},\;1,\;10^3$, respectively. Dashed lines indicate
    mean protein level and mean protein level plus standard deviation.
    Black lines correspond to the high trigger level and spiking
    events are indicated by black circles. (D, E, F) Probabilities of
    observing $n$ spikes during a time window of 10 averaged
    transition $\rho=10^{-3},\;1,\;10^3$, respectively.

\subsubsection*{Figure~\ref{fig:fig5}.}
\textbf{Regularity of stochastic oscillations} In (A), (B),
    and (C), the value of the Fano factor $F$, which quantifies
    spiking regularity, is shown as a function of two parameters using
    a gray-scale color code with level lines displayed in red. The
    white lines in (A) and (B) enclose the region where the reduced
    model~(\ref{eq:Mod_red}), discussed in
    Sec.~\ref{sec:determ-model-oscillations}, predicts oscillations.
    The regularity of stochastic oscillations is favored by balanced
    protein and mRNA degradation rates (corresponding to $\delta\simeq
    1$) as well as (A) a resonance parameter $\rho$ close to 1, (B) a
    high value of the overall production rate $\Lambda$. (C) The
    ratio $\beta/\delta$ controlling the relative mRNA to protein
    concentration has no effect on spike regularity. (D) The
    oscillation period (or protein average interspike time interval)
    is controlled by the resonance parameter $\rho$. The variation of
    $\chi$, the average number of ``on''/ ``off'' cycles in an
    interspike interval, is displayed using a gray color code as a
    function of the lifetime ratio $\delta$ and of the resonance
    parameter $\rho$. Level lines are displayed in red (color online).
    Stochastic simulations of the biochemical network have been
    carried out with $k_{off}/k_{on}=100$ ; $\beta=10\delta$ (A, B,
    and D); $\rho=1$ (B and C); $\Lambda=100$ (A, B, and C ).

\clearpage

\clearpage

\begin{figure}[htbp]
  \begin{center}
    \includegraphics*{Wang_Stochastic_Oscillations2}
    \caption[]{}
    \label{fig:fig2}
  \end{center}
\end{figure}

\clearpage

\begin{figure}[htbp]
  \begin{center}
    \includegraphics*{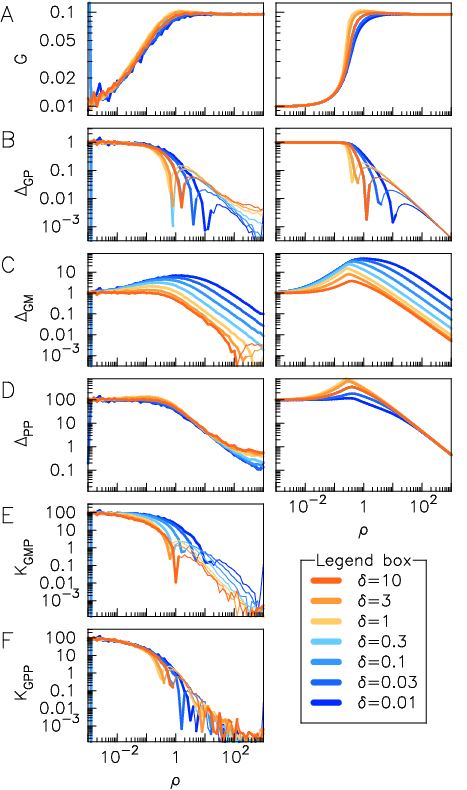}
    \caption[]{}
    \label{fig:fig3}
  \end{center}
\end{figure}

\clearpage

\begin{figure}[htbp]
  \begin{center}
    \includegraphics*[width=6in]{Wang_Stochastic_Oscillations4}
    \caption[]{}
    \label{fig:fig4}
  \end{center}
\end{figure}

\clearpage

\begin{figure}[htbp]
  \begin{center}
    \includegraphics*[width=6in]{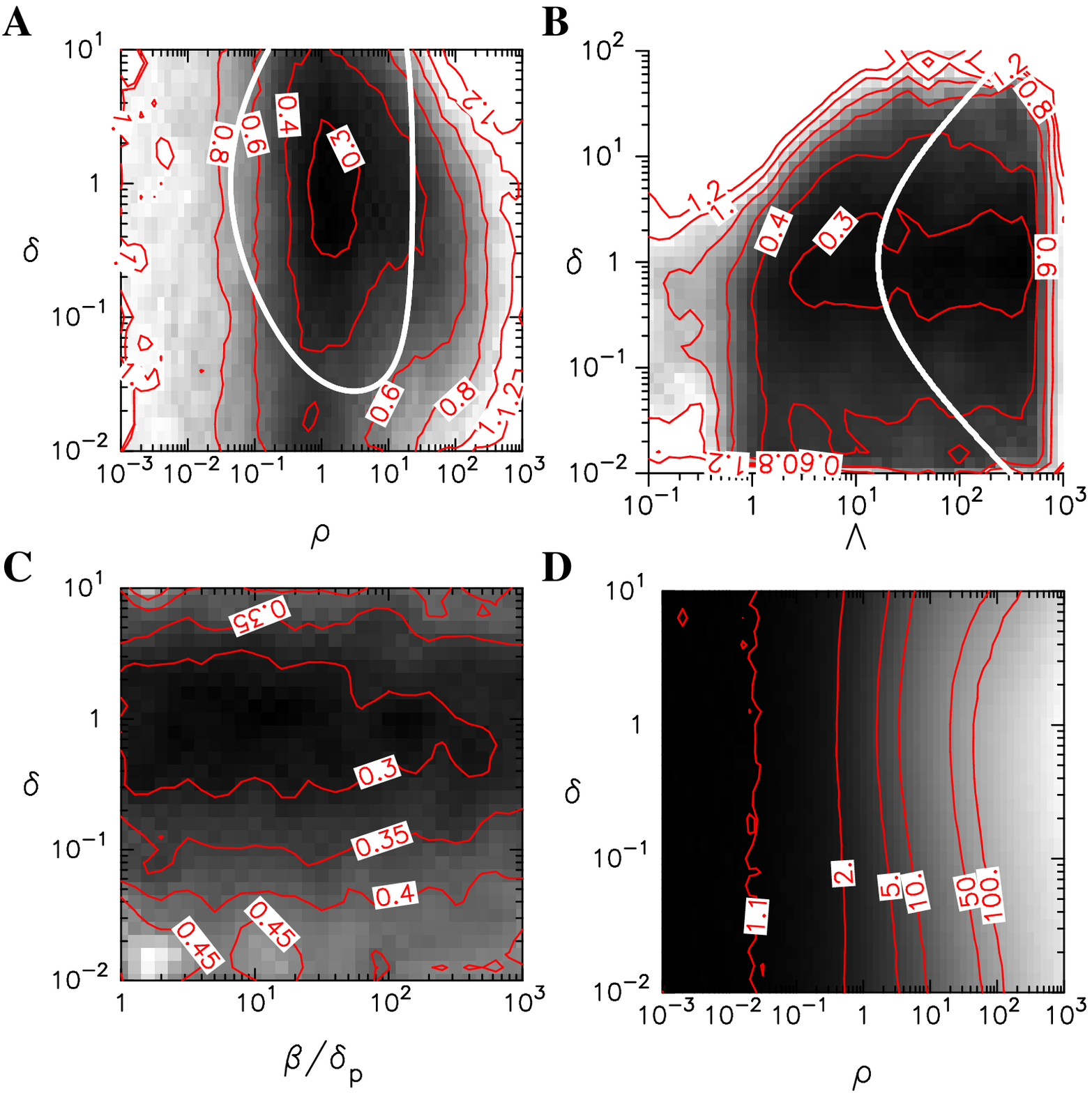}
    \caption[]{}
    \label{fig:fig5}
  \end{center}
\end{figure}


\begin{thebibliography}{54}
\providecommand{\url}[1]{\texttt{#1}}
\providecommand{\urlprefix}{ }

\bibitem[Hartwell et~al.(1999)Hartwell, Hopfield, Leibler, and
  Murray]{Hartwell99}
Hartwell, L., J.~Hopfield, S.~Leibler, and A.~Murray, 1999.
\newblock From molecular to modular cell biology.
\newblock \emph{Nature} 402:C47--C52.

\bibitem[Golding et~al.(2005)Golding, Paulsson, Zawilski, and Cox]{Golding05}
Golding, I., J.~Paulsson, S.~M. Zawilski, and E.~C. Cox, 2005.
\newblock Real-time kinetics of Gene activity in Individual bacteria.
\newblock \emph{Cell} 113:1025--1036.

\bibitem[Darzacq et~al.(2007)Darzacq, Shav-Tal, de~Turris, Brody, Shenoy,
  Phair, and Singer]{xavier07}
Darzacq, X., Y.~Shav-Tal, V.~de~Turris, Y.~Brody, S.~M. Shenoy, R.~D. Phair,
  and R.~H. Singer, 2007.
\newblock In vivo dynamics of RNA polymerase II transcription.
\newblock \emph{Nat. Struct. Mol. Biol.} 14:796--806.

\bibitem[Chubb et~al.(2006)Chubb, Trcek, Shenoy, and Singer]{Chubb06}
Chubb, J.~R., T.~Trcek, S.~M. Shenoy, and R.~H. Singer, 2006.
\newblock Transcriptional Pulsing of a Developmental Gene.
\newblock \emph{Curr. Biol.} 16:1018--1025.

\bibitem[Suter et~al.(2011)Suter, Molina, Gatfield, Schneider, Schibler, and
  Naef]{Suter11:_mammalian_bursting}
Suter, D.~M., N.~Molina, D.~Gatfield, K.~Schneider, U.~Schibler, and F.~Naef,
  2011.
\newblock Mammalian Genes Are Transcribed with Widely Different Bursting
  Kinetics.
\newblock \emph{Science} 332:472--474.
\newblock
  \urlprefix\url{http://www.sciencemag.org/content/332/6028/472.abstract}.

\bibitem[Harper et~al.(2011)Harper, Finkenstädt, Woodcock, Friedrichsen,
  Semprini, Ashall, Spiller, Mullins, Rand, Davis, and
  White]{Harper11:_stochastic_cycles}
Harper, C.~V., B.~Finkenstädt, D.~J. Woodcock, S.~Friedrichsen, S.~Semprini,
  L.~Ashall, D.~G. Spiller, J.~J. Mullins, D.~A. Rand, J.~R.~E. Davis, and
  M.~R.~H. White, 2011.
\newblock Dynamic Analysis of Stochastic Transcription Cycles.
\newblock \emph{PLoS Biol} 9:e1000607.
\newblock \urlprefix\url{http://dx.doi.org/10.1371\%2Fjournal.pbio.1000607}.

\bibitem[Eldar and Elowitz(2010)]{eldar10:_funct}
Eldar, A., and M.~Elowitz, 2010.
\newblock Functional roles for noise in genetic circuits.
\newblock \emph{Nature} 467:167--273.

\bibitem[Simpson et~al.(2003)Simpson, Cox, and Sayler]{Simpson15042003}
Simpson, M.~L., C.~D. Cox, and G.~S. Sayler, 2003.
\newblock Frequency domain analysis of noise in autoregulated gene circuits.
\newblock \emph{Proceedings of the National Academy of Sciences}
  100:4551--4556.
\newblock \urlprefix\url{http://www.pnas.org/content/100/8/4551.abstract}.

\bibitem[Warren et~al.(2006)Warren, Tanase-Nicola, and ten
  Wolde]{warren:144904}
Warren, P.~B., S.~Tanase-Nicola, and P.~R. ten Wolde, 2006.
\newblock Exact results for noise power spectra in linear biochemical reaction
  ne tworks.
\newblock \emph{The Journal of Chemical Physics} 125:144904.
\newblock \urlprefix\url{http://link.aip.org/link/?JCP/125/144904/1}.

\bibitem[Lestas et~al.(2008)Lestas, Paulsson, Ross, and
  Vinnicombe]{lestas2008noise}
Lestas, I., J.~Paulsson, N.~Ross, and G.~Vinnicombe, 2008.
\newblock Noise in gene regulatory networks.
\newblock \emph{Automatic Control, IEEE Transactions on} 53:189--200.

\bibitem[Aquino et~al.(2012)Aquino, Abranches, and Nunes]{Aquino12}
Aquino, T., E.~Abranches, and A.~Nunes, 2012.
\newblock Stochastic single-gene autoregulation.
\newblock \emph{Phys. Rev. E} 85:061913.
\newblock \urlprefix\url{http://link.aps.org/doi/10.1103/PhysRevE.85.061913}.

\bibitem[Stricker et~al.(2008)Stricker, Cookson, Bennett, Mather, Tsimring, and
  Hasty]{Stricker08:robust_oscillator}
Stricker, J., S.~Cookson, M.~R. Bennett, W.~H. Mather, L.~S. Tsimring, and
  J.~Hasty, 2008.
\newblock A fast, robust and tunable synthetic gene oscillator.
\newblock \emph{Nature} 456:516--520.

\bibitem[Hermsen et~al.(2010)Hermsen, Ursem, and ten Wolde]{Hermsen2010a}
Hermsen, R., B.~Ursem, and P.~R. ten Wolde, 2010.
\newblock Combinatorial Gene Regulation Using Auto-Regulation.
\newblock \emph{PLoS Comput Biol} 6:e1000813.
\newblock \urlprefix\url{http://dx.doi.org/10.13712Fjournal.pcbi.1000813}.

\bibitem[Salgado et~al.(2001)Salgado, Santos-Zavaleta, Gama-Castro,
  Millán-Zárate, Díaz-Peredo, Sánchez-Solano, Pérez-Rueda, Bonavides-Martínez,
  and Collado-Vides]{Salgado01012001}
Salgado, H., A.~Santos-Zavaleta, S.~Gama-Castro, D.~Millán-Zárate,
  E.~Díaz-Peredo, F.~Sánchez-Solano, E.~Pérez-Rueda, C.~Bonavides-Martínez, and
  J.~Collado-Vides, 2001.
\newblock RegulonDB (version 3.2): transcriptional regulation and operon
  organization in Escherichia coli K-12.
\newblock \emph{Nucleic Acids Research} 29:72--74.
\newblock
  \urlprefix\url{http://nar.oxfordjournals.org/content/29/1/72.abstract}.

\bibitem[Keseler et~al.(2005)Keseler, Collado-Vides, Gama-Castro, Ingraham,
  Paley, Paulsen, Peralta-Gil, and Karp]{Keseler01012005}
Keseler, I.~M., J.~Collado-Vides, S.~Gama-Castro, J.~Ingraham, S.~Paley, I.~T.
  Paulsen, M.~Peralta-Gil, and P.~D. Karp, 2005.
\newblock EcoCyc: a comprehensive database resource for Escherichia coli.
\newblock \emph{Nucleic Acids Research} 33:D334--D337.
\newblock
  \urlprefix\url{http://nar.oxfordjournals.org/content/33/suppl\_1/D334.abstra%
ct}.

\bibitem[Novak and Tyson({2008})]{Novak_08}
Novak, B., and J.~J. Tyson, {2008}.
\newblock {Design principles of biochemical oscillators}.
\newblock \emph{{Nat. Rev. Mol. Cell Biol.}} {9}:{981--991}.

\bibitem[Hirata et~al.(2002)Hirata, Yoshiura, Ohtsuka, Bessho, Harada,
  Yoshikawa, and Kageyama]{Hirata02a}
Hirata, H., S.~Yoshiura, T.~Ohtsuka, Y.~Bessho, T.~Harada, K.~Yoshikawa, and
  R.~Kageyama, 2002.
\newblock Oscillatory expression of the b{HLH} factor {H}es1 regulated by a
  negative feedback loop.
\newblock \emph{Science} 298:840--843.

\bibitem[Goodwin(1965)]{Goodwin65a}
Goodwin, B.~C., 1965.
\newblock Oscillatory behavior of enzymatic control processes.
\newblock \emph{Adv. Enzyme Regul.} 3:425--439.

\bibitem[Griffith(1968)]{Griffith68a}
Griffith, J.~S., 1968.
\newblock Mathematics of cellular control processes I. Negative feedback to one
  gene.
\newblock \emph{J. Theor. Biol.} 20:202--208.

\bibitem[Bliss et~al.(1982)Bliss, Painter, and Marr]{Bliss82a}
Bliss, R.~D., P.~R. Painter, and A.~G. Marr, 1982.
\newblock Role of feedback inhibition in stabilizing the classical operon.
\newblock \emph{J. Theor. Biol.} 97:177--193.

\bibitem[Goldbeter(1995)]{Goldbeter95a}
Goldbeter, A., 1995.
\newblock A model for circadian oscillations in the {D}rosophila period protein
  (PER).
\newblock \emph{Proc. R. Soc. Lond. B} 261:319--324.

\bibitem[Leloup et~al.(1999)Leloup, Gonze, and Goldbeter]{Leloup99a}
Leloup, J.-C., D.~Gonze, and A.~Goldbeter, 1999.
\newblock Limit cycle models for circadian rhythms based on transcriptional
  regulation in {D}rosophila and {N}eurospora.
\newblock \emph{J. Biol. Rhythms} 14:433--448.

\bibitem[Lewis(2003)]{Lewis2003a}
Lewis, J., 2003.
\newblock Autoinhibition with transcriptional delay: a simple mechanism for the
  zebrafish somitogenesis oscillator.
\newblock \emph{Curr. Biol.} 13:1398--1408.

\bibitem[Monk(2003)]{Monk2003a}
Monk, N.~A.~M., 2003.
\newblock Oscilatory expression of Hes1, p53 and NK-$\kappa$B driven by
  transcriptional time delays.
\newblock \emph{Curr. Biol.} 13:1409--1413.

\bibitem[Jensen et~al.(2003)Jensen, Sneppen, and Tiana]{Jensen2003a}
Jensen, M.~H., K.~Sneppen, and G.~Tiana, 2003.
\newblock Sustained oscillations and time delays in gene expression of protein
  Hes1.
\newblock \emph{FEBS Lett.} 541:176--177.

\bibitem[Morant et~al.(2009)Morant, Thommen, Lemaire, Vandermo\"ere, Parent,
  and Lefranc]{Morant09}
Morant, P.-E., Q.~Thommen, F.~Lemaire, C.~Vandermo\"ere, B.~Parent, and
  M.~Lefranc, 2009.
\newblock Oscillations in the Expression of a Self-Repressed Gene Induced by a
  Slow Transcriptional Dynamics.
\newblock \emph{Phys. Rev. Lett.} 102:068104.
\newblock
  \urlprefix\url{http://link.aps.org/doi/10.1103/PhysRevLett.102.068104}.

\bibitem[Tiana et~al.(2007)Tiana, Krishna, Pigolotti, Jensen, and
  Sneppen]{tiana07:_oscil}
Tiana, G., S.~Krishna, S.~Pigolotti, M.~H. Jensen, and K.~Sneppen, 2007.
\newblock Oscillations and temporal signalling in cells.
\newblock \emph{Phys. Biol.} 4:R1--R17.

\bibitem[Mengel et~al.(2010)Mengel, Hunziker, Pedersen, Trusina, Jensen, and
  Krishna]{Jensen_Current_Opinion}
Mengel, B., A.~Hunziker, L.~Pedersen, A.~Trusina, M.~H. Jensen, and S.~Krishna,
  2010.
\newblock Modelling oscillatory control in NF-kB, p53, and Wnt signaling.
\newblock \emph{Current Opinion in Genetics and Development} 20:656--664.

\bibitem[Tyson et~al.(1999)Tyson, Hong, D.Thron, and Novak]{Tyson99:_per_tim}
Tyson, J.~J., C.~I. Hong, C.~D.Thron, and B.~Novak, 1999.
\newblock A simple model of circadian rhythms based on dimerization and
  proteolysis of PER and TIM.
\newblock \emph{Biophys. J.} 77:2411--2417.

\bibitem[Fran\c{c}ois and Hakim(2005)]{Francois05}
Fran\c{c}ois, P., and V.~Hakim, 2005.
\newblock Core genetic module: the mixed feedback loop.
\newblock \emph{Phys. Rev. E} 72:031908.

\bibitem[van Kampen(2007)]{VanKampen}
van Kampen, N.~G., 2007.
\newblock Stochastic processes in physics and chemistry.
\newblock Elsevier.

\bibitem[Hornos et~al.(2005)Hornos, Schultz, Innocentini, Wang, Walczak,
  Onuchic, and Wolynes]{hornos05}
Hornos, J.~E.~M., D.~Schultz, G.~C.~P. Innocentini, J.~Wang, A.~M. Walczak,
  J.~N. Onuchic, and P.~G. Wolynes, 2005.
\newblock Self-regulating gene: An exact solution.
\newblock \emph{Phys. Rev. E} 72:051907.

\bibitem[Grima et~al.(2012)Grima, Schmidt, and Newman]{grima2012}
Grima, R., D.~Schmidt, and T.~Newman, 2012.
\newblock Steady-state fluctuations of a genetic feedback loop: An exact
  solution.
\newblock \emph{J. Chem. Phys.} 137:035104.

\bibitem[Elf and Ehrenberg(2003)]{elf03:_lna}
Elf, J., and M.~Ehrenberg, 2003.
\newblock Fast evaluation of fluctuations in biochemical networks with the
  linear noise approximation.
\newblock \emph{Genome Res.} 13:2475--2484.

\bibitem[McKane and Newman(2005)]{mckane05:_predat}
McKane, A.~J., and T.~J. Newman, 2005.
\newblock Predator-prey cycles from resonant amplification of demographic
  stochasticity.
\newblock \emph{Phys. Rev. Lett.} 94:218102.

\bibitem[McKane et~al.(2007)McKane, Nagy, Newman, and
  Stefanini]{mckane07:_amplif}
McKane, A.~J., J.~D. Nagy, T.~J. Newman, and M.~O. Stefanini, 2007.
\newblock Amplified biochemical oscillations in cellular systems.
\newblock \emph{J. Stat. Phys.} 128:165--191.

\bibitem[Galla({2009})]{Galla_09}
Galla, T., {2009}.
\newblock {Intrinsic fluctuations in stochastic delay systems: Theoretical
  description and application to a simple model of gene regulation}.
\newblock \emph{{Phys. Rev. E}} {80}:{021909}.

\bibitem[Loinger and Biham(2007)]{loinger:051917}
Loinger, A., and O.~Biham, 2007.
\newblock Stochastic simulations of the repressilator circuit.
\newblock \emph{Phys. Rev. E} 76:051917.

\bibitem[Blossey et~al.(2008)Blossey, Cardelli, and Phillips]{blossey:2008}
Blossey, R., L.~Cardelli, and A.~Phillips, 2008.
\newblock Compositionality, stochasticity, and cooperativity in dynamic models
  of gene regulation.
\newblock \emph{HFSP Journal} 2:17--28.

\bibitem[Barrio et~al.({2006})Barrio, Burrage, Leier, and Tian]{Barrio_06}
Barrio, M., K.~Burrage, A.~Leier, and T.~Tian, {2006}.
\newblock {Oscillatory regulation of hes1: Discrete stochastic delay modelling
  and simulation}.
\newblock \emph{{PLoS Comput. Biol.}} {2}:{1017--1030}.

\bibitem[Lepzelter et~al.(2010)Lepzelter, Feng, and Wang]{Lepzelter_10}
Lepzelter, D., H.~Feng, and J.~Wang, 2010.
\newblock Oscillation, cooperativity, and intermediates in the self-repressing
  gene.
\newblock \emph{Chem. Phys. Lett.} 490:216.

\bibitem[Scott et~al.({2007})Scott, Hwa, and Ingalls]{Scott07}
Scott, M., T.~Hwa, and B.~Ingalls, {2007}.
\newblock {Deterministic characterization of stochastic genetic circuits}.
\newblock \emph{{Proc. Nat. Acad. Sci. USA}} {104}:{7402--7407}.

\bibitem[Gillespie(2009)]{gillespie09:_moment}
Gillespie, C.~S., 2009.
\newblock Moment-closure approximations for mass action models.
\newblock \emph{IET Syst. Biol.} 3:52--58.

\bibitem[Lee et~al.(2009)Lee, Kim, and Kim]{lee09:closure}
Lee, C.~H., K.-H. Kim, and P.~Kim, 2009.
\newblock A moment closure method for stochastic reaction networks.
\newblock \emph{J. Chem. Phys.} 130:134107.
\newblock \urlprefix\url{http://link.aip.org/link/?JCP/130/134107/1}.

\bibitem[Lafuerza and Toral(2010)]{lafuerza10:_gauss}
Lafuerza, L.~F., and R.~Toral, 2010.
\newblock On the {G}aussian approximation for Master Equations.
\newblock \emph{J. Stat. Phys.} 140:917--933.

\bibitem[Coulon et~al.(2010)Coulon, Gandrillon, and Beslon]{coulon2010}
Coulon, A., O.~Gandrillon, and G.~Beslon, 2010.
\newblock On the spontaneous stochastic dynamics of a single gene: complexity
  of the molecular interplay at the promoter.
\newblock \emph{BMC systems biology} 4:2.

\bibitem[Fano({1947})]{Fano47}
Fano, U., {1947}.
\newblock {Ionization yield of radiation .2. The fluctuations of the number of
  ions}.
\newblock \emph{{Phys. Rev.}} {72}:{26--29}.

\bibitem[Bernard et~al.(2006)Bernard, Cajavec, Pujo-Menjouet, Mackey, and
  Herzel]{bernard06:_hes1}
Bernard, S., B.~Cajavec, L.~Pujo-Menjouet, M.~Mackey, and H.~Herzel, 2006.
\newblock Modelling transcriptional feedback loops: the role of Gro/TLE1 in
  Hes1 oscillations.
\newblock \emph{Phil. Trans. R. Soc. A} 364:1155--1170.

\bibitem[Bratsun et~al.(2005)Bratsun, Volfson, Tsimring, and
  Hasty]{Bratsun05:delay_stochastic_oscillations}
Bratsun, D., D.~Volfson, L.~S. Tsimring, and J.~Hasty, 2005.
\newblock Delay-induced stochastic oscillations in gene regulation.
\newblock \emph{Proc. Nat. Acad. Sci. USA} 102:14593--14598.
\newblock \urlprefix\url{http://www.pnas.org/content/102/41/14593.abstract}.

\bibitem[Song et~al.(2010)Song, Phenix, Abedi, Scott, Ingalls, Kærn, and
  Perkins]{Song10:_stochastic_bifurcation}
Song, C., H.~Phenix, V.~Abedi, M.~Scott, B.~P. Ingalls, M.~Kærn, and T.~J.
  Perkins, 2010.
\newblock Estimating the Stochastic Bifurcation Structure of Cellular Networks.
\newblock \emph{PLoS Comput. Biol.} 6:e1000699.
\newblock \urlprefix\url{http://dx.doi.org/10.1371\%2Fjournal.pcbi.1000699}.

\bibitem[Vidal et~al.(2012)Vidal, Petitot, Boulier, Lemaire, and
  Kuttler]{Vidal2012}
Vidal, S., M.~Petitot, F.~Boulier, F.~Lemaire, and C.~Kuttler, 2012.
\newblock Models of Stochastic Gene Expression and Weyl Algebra.
\newblock \emph{In} K.~Horimoto, M.~Nakatsui, and N.~Popov, editors, Algebraic
  and Numeric Biology, Springer Berlin Heidelberg, volume 6479 of \emph{Lecture
  Notes in Computer Science}, 76--97.
\newblock \urlprefix\url{http://dx.doi.org/10.1007/978-3-642-28067-2\_5}.

\bibitem[Purcell et~al.(2010)Purcell, Savery, Grierson, and
  di~Bernardo]{Purcell06112010}
Purcell, O., N.~J. Savery, C.~S. Grierson, and M.~di~Bernardo, 2010.
\newblock A comparative analysis of synthetic genetic oscillators.
\newblock \emph{J. R. Soc. Interface} 7:1503--1524.
\newblock
  \urlprefix\url{http://rsif.royalsocietypublishing.org/content/7/52/1503.abst%
ract}.

\bibitem[Murugan and Kreiman({2011})]{Murugan11}
Murugan, R., and G.~Kreiman, {2011}.
\newblock {On the Minimization of Fluctuations in the Response Times of
  Autoregulatory Gene Networks}.
\newblock \emph{{Biophys. J.}} {101}:{1297--1306}.

\bibitem[Gibson and Bruck(2000)]{GibsonBruck00}
Gibson, M.~A., and J.~Bruck, 2000.
\newblock Efficient Exact Stochastic Simulation of Chemical Systems with Many
  Species and Many Channels.
\newblock \emph{J. Phys. Chem. A} 104:1876--1889.

\end{thebibliography}
\end{document}